# SNE IA: ON THE BINARY PROGENITORS AND EXPECTED STATISTICS

P. RUIZ–LAPUENTE

*Department of Astronomy, University of Barcelona*
*Martí i Franqués, 1 – 08028 Barcelona, Spain*

*Max–Planck–Institut für Astrophysik*
*Karl–Schwarzschildstrasse 1, D–69117 Garching bei München, Germany*

R. CANAL

*Department of Astronomy, University of Barcelona*
*Martí i Franqués, 1 – 08028 Barcelona, Spain*

AND

A. BURKERT

*Max–Planck–Institut für Astronomie*
*Königstuhl 17, D–69117 Heidelberg, Germany*

## 1. Introduction

Type Ia supernovae (SNe Ia) are explosions of C+O WDs triggered by the accretion of material from a companion. After many years of SN research, we are fairly certain of the nature of the exploding object, but very little is still known about the companion star. In fact, it has been from long noticed that the companion does not leave its imprint on the explosion in a way which could allow us to identify the progenitor system. It is hoped that in the near future the various observational tests discussed along this conference will shed light into the identification issue.

One of the approaches to evaluate the possible binary scenarios leading to SNe Ia is the study of rates and age contraints on the exploding system. SN statistics is a rapidly developing field. In the particular case of SNe Ia, it includes not only the determination of the SNe Ia rates in galaxies of different types (see Cappellaro & Turatto, this volume) but also that of the possible systematic variation along the Hubble sequence of average SNe Ia characteristics such as peak luminosity and expansion velocity of the ejecta (Branch, this volume), light–curve shape (Phillips 1993), or the amplitude of dispersion in any of those properties (Filippenko 1989, and this volume). Besides, the ongoing searches for SNe Ia at large redshifts are already providing the means to link SNe Ia statistics with cosmic evolution (see Pain et al., Perlmutter et al., and Schmidt, all in this volume).

On the other side, for any given combination of SN Ia scenario plus explosion model, we can predict the corresponding SN Ia rates and average SNe Ia characteristics. Whelan & Iben (1973), presenting the first detailed scenario for SNe Ia, included already an estimate of the current galactic SNe Ia rate to be expected from it. Later, Iben & Tutukov (1984), examining different SNe Ia scenarios, also predicted the SNe Ia rates for our Galaxy at the present time, and used the comparison with the observational estimate to select the most plausible scenarios. This kind of test has been systematically applied thereafter to every proposed SN Ia scenario by different groups (Tutukov & Yungelson 1994; Branch et al. 1995; Ruiz–Lapuente, Burkert & Canal 1995; Canal, Ruiz–Lapuente & Burkert 1996; Di Stefano et al., this volume). This sort of studies rely on results obtained from detailed evolutionary calculations (see, for instance Iben 1992; Hernanz et al, this volume). The possibility of edge–lit detonations of WDs below the Chandrasekhar mass (sub–Chandrasekhar explosions, hereafter sub–Ch) in addition to that of central ignition of a WD reaching the Chandrasekhar–mass (Ch explosions) have been considered in our own work, and we have simulated the characteristics of the resulting SNe Ia for the different galaxy types (Ruiz–Lapuente, Burkert & Canal 1995, thereafter RBC95, and Canal, Ruiz–Lapuente & Burkert 1996, thereafter CRB96). The dependence of SNe Ia rates and the aforementioned characteristics on cosmological redshift $z$ has also been derived (RBC95), and this could provide a new means to discriminate among scenarios as data from ever more complete SNe Ia surveys, extending to ever larger redshifts, become available. Conversely, given the right combination of SNe Ia scenario plus explosion model, SNe Ia statistics can be used to reconstruct the star–formation histories of galaxies along cosmic evolution.

A current major goal of our modeling of SNe Ia statistics is to decide whether a single scenario plus a single explosion model can explain all the observed trends or, on the contrary, more than one scenario (and/or more than one explosion model) is required. Next section summarizes the main observed statistical trends which should be reproduced by the sort of Monte Carlo calculations done in this work. Afterwards we describe the main SNe Ia scenarios proposed so far, and we present the numerical approximations to the results from detailed calculations of close binary evolution which are used.

## 2. Observed Statistics

The statistical correlations of the characteristics of the explosions (maximum brightness, expansion velocities of the ejecta, and rate of decline of the light curve after maximum) with their sites provide very important pieces of information to elucidate which systems are the progenitors of SNe Ia. The ultimate goal of such studies is to infer the age of the populations involved in the explosions and to explore the variation of the SNe Ia rates (absolute SNe Ia rates and SNe Ia/SNe II proportions) as functions of galaxy type and of redshift.

Observed rates of SNe are normally given in frequency ($yr^{-1}$) per unit blue luminosity ($10^{10}$ $L_{B\odot}$). The standard unit is called SNU, one SNU corresponding to 1 SN per 100 years per $10^{10}$ $L_{B\odot}$ (Tammann 1994). Since the luminosity attributed to the galaxy depends on the Hubble constant $H_0$, there is a factor $h^2$ ($h \equiv H_0/50\ km\ s^{-1}\ Mpc^{-1}$) involved in the extragalactic SN rates.

SNe Ia, in contrast with the other supernova types (SNe II, SNe Ib/c), appear in all types of galaxies, but their rates depend on the Hubble type of the parent galaxy. Cappellaro et al. (1993) find that the SNe Ia rates (in SNU) are 3 times higher in late–type (Sbc–Sd) galaxies than in early–type (E–S0) galaxies (see also Cappellaro & Turatto, this volume). Della Valle & Livio (1994) find an even more marked contrast: the SNe Ia rate (per unit of K luminosity of the parent galaxy) would be 5–10 times larger in late spirals than in ellipticals. This trend is also found in the numerical simulations presented below: the higher frequencies of SNe Ia in late–type galaxies, as compared with early–type ones, reflect the fact that star formation is still going on in spirals and irregulars whilst it stopped several Gyr ago in ellipticals. This effect is significantly smeared out if the rates are given relative to blue luminosity $L_B$, due to the fact that the ratio $M_{lum}/L_B(\odot)$ decreases along the Hubble sequence. According to Bartunov & Tsvetkov (this meeting) rates of SNe Ia in spiral galaxies are higher in the neighborhood of the spiral arms.

It has been shown (Hamuy et al. 1995; Phillips and Suntzeff, both in this meeting) that the absolute brightness of SNe Ia at maximum correlates with the morphological type of the parent galaxy: on average, SNe Ia are brighter in late–type (spiral and irregular) than in early–type (elliptical and lenticular) galaxies (Branch, this volume; Hamuy et al. 1995; Branch, Romanishin, & Baron 1996; van den Bergh 1996). The amplitude of the variation is under discussion. Branch, Romanishin, & Baron (1996) find that when very subluminous and spectroscopically peculiar SNe Ia are dis-regarded, SNe Ia in red (early–type) galaxies are less luminous than those in blue (late–type) galaxies by 0.3 mag. van den Bergh (1996) finds the same effect, although he favors a broader range.

At the same time, a correlation has now been well established, basically from the extensive and very precise photometric observations made at CTIO (Phillips 1993; Maza et al. 1994; Hamuy et al. 1995), that brighter SNe Ia decline more slowly after maximum than fainter ones. This had earlier been quantified by the speed–class parameter $\beta$ (Pskovskii 1984). Phillips (1993) uses $\Delta m_{15}$, the decrease in B–magnitude between maximum and 15 days past maximum, whereas consideration of the light curve shape as a whole has been introduced by Riess, Press, & Kirshner (1995) (see also Vacca & Leibundgut, this volume).

On the other hand, Filippenko (1989) noticed that the dispersion in the speed class of the light curves of SNe Ia (then measured by the parameter $\beta$) increases when going from early towards late–type galaxies: the range of $\beta$–values ($\Delta m_{15}$ in the modern use) is smaller in ellipticals than in the other types of galaxies (see also Filippenko, this volume). That coincides with the broader dispersion of supernova characteristics in spiral galaxies as compared with ellipticals predicted by models (RBC95; CRB96), which corresponds to the equally broader dispersion in the ages of the progenitors.

The expansion velocities of the ejecta, as measured from the blueshifts of the characteristic Si II $\lambda 6335$ absorption line at maximum light, also correlate with the Hubble type of the parent galaxy: the average velocity increases along the Hubble sequence from E0 to Sd, the amplitude of the variation being $\sim 5000$ km s$^{-1}$ (Branch & van den Bergh 1993; Branch, this volume). The dispersion in the blueshifts seems to increase from early to late–type galaxies (Branch & van den Bergh 1993), as it occurs with the light–curve shape parameter $\Delta m_{15}$ (or $\beta$).

## 3. Overview of SNe Ia binary progenitors

We will consider five different types of binary systems as possible SNe Ia progenitors, according to the companion of the exploding WD. Such elusive companion might be another C+O WD, which is a particular case of *double degenerate* (DD) *system*. It might be a He star (star presently burning He at is center, but having been depleted of its H envelope in a previous common envelope episode). In that case, we refer to the system as a *He–star cataclysmic* (HeCV). The companion might be a main-sequence (MS) or subgiant (SG) star of higher mass than the WD which is filling its Roche lobe and transfers mass on a thermal timescale. Such system is termed by us as a *cataclysmic–like system* (CLS). It is also been called "H–Algol" in the context of SNe Ia (Branch et al. 1995). The donor star might be an asymptotic giant branch (AGB) star emitting a strong wind which provides the H to be accreted, thus forming a *symbiotic system* (SS). A further H donor might be a MS star or SG which fills its Roche lobe due to the magnetic braking mechanism (Verbunt & Zwaan 1981), and we refer to the corresponding systems as cataclysmic variables (CV). In scenarios based on DD systems, only Ch WD explosions are possible, since the accretion of C+O only produces explosion when the WD is close to the Chandrasekhar limit. Scenarios involving CLS or CV, instead, allow both central ignition of Ch C+O WDs or edge–lit detonations of sub–Ch WDs, by the conversion of the accreted H into He, and the burning of He into C+O either in non–degenerate (first case) or degenerate conditions (second case). In HeCV and SS only sub–Ch explosions are likely.

### 3.1. DOUBLE DEGENERATES

The lack of discovery of double degenerate C+O systems able to merge in less than a Hubble time (Bragaglia et al., this volume) would apparently disfavor DD systems as possible SNe Ia progenitors. However, Isern et al. (this volume) find that if the cooling process of the WD is duly taken into account, the negative results of the spectroscopic searches do not invalidate DDs as posible SNe Ia progenitors. More serious objections to this scenario come from hydrodynamic arguments (Mochkovitch, Guerrero & Segretain, this volume).

In the DD scenario (Iben & Tutukov 1984; Iben, this volume), an initially detached binary consisting of two intermediate–mass MS stars undergoes two successive common–envelope (CE) episodes when the primary and the secondary, by turn, fill their Roche lobes during the AGB phase. Each time, the CE is expelled and a C+O WD forms from the core of the AGB star. Ejection of the envelope results from the conversion of orbital energy into internal energy of the envelope during spiral–in of the less massive component. The separation $A$ of the system is reduced from its primordial value ($A_0$) by an amount which depends on the poorly known *common–envelope parameter,* $\alpha$ (see below). Afterwards, the orbit further shrinks due to the emission of gravitational–wave radiation (GWR), until the less massive WD overfills its Roche lobe. A runaway mass transfer process starts (the radius of a WD increases when its mass decreases), which leads to the formation of a massive disk or torus around the more massive WD, coming from the disruption (on a quasidynamic time scale) of the companion (Benz et al. 1990). Subsequent accretion of this material would bring the remaining WD to the point of explosive C ignition at its center and produce a SN Ia. Depending on the combination of initial masses and primordial separations, DD pairs made of a C+O WD plus an O+Ne+Mg or a He WD might also form, or still a pair of He WDs or two O+Ne+Mg WDs. For reasons to be given below, we only consider C+O WD pairs as possible progenitors of SNe Ia.

The primary stars involved in this scenario are in the initial mass range $5M_\odot \lesssim M_1 \lesssim 8M_\odot$. The masses of the two components of the binary system should be similar, which we have introduced in our modeling by the condition $0.5 \leq q \leq 1$ ($q \equiv M_2/M_1$, $M_2$ being the mass of the secondary). The initial separations must be large enough to allow formation of the two C+O WDs prior to merging and small enough for CE episodes to be possible. That translates into the condition: $70R_\odot \lesssim A_0 \lesssim 1500R_\odot$. The masses of the two WDs can be related to the initial masses of the primary and the secondary by the relationship (Iben & Tutukov 1984):

$$M_{WD1,2} \simeq 0.1 \times M_{1,2}^{1.2} \tag{1}$$

which approximately reproduces the results of van den Linden (1980) for the core mass *vs* total mass relationship in AGB stars. We have compared the results from (1) with the use of the star radius–core mass diagram of Politano (1988) and de Kool (1992) and found that the differences are not significant.

An additional constraint, in this scenario, is that $M_{WD1} + M_{WD2} \gtrsim 1.4 M_\odot$. The final separation after the two CE episodes goes as:

$$A_{ff} = \alpha^2 \left(\frac{M_{WD1}}{M_1}\right)^2 \frac{M_{WD2}}{M_2} A_0 \tag{2}$$

The CE parameter, $\alpha$, is a measure of the efficiency with which orbital energy is used in the ejection of the common envelope. We have adopted the definition by Tutukov & Yungelson (1979):

$$\alpha \frac{GM_c M_2}{A_f} = \frac{GM_1^2}{A_0} \tag{3}$$

where $M_c$ is the mass of the degenerate C+O core of the primary (or the mass of its He core, if Roche lobe overflow occurs at an earlier stage).

The value of $\alpha$ is uncertain. Hydrodynamic simulations by Livio & Soker (1988) and by Taam & Bodenheimer (1989) indicate $\alpha \simeq 0.3 - 0.6$, but arguments for higher values are given by Yungelson et al. (1994) and Tutukov & Yungelson (1994). Values $\alpha = 1$ and $\alpha = 0.5$ have been used in the simulations of SNe Ia statistics of RBC95, and $\alpha = 1$ and $\alpha = 0.3$ in those of CRB96. The explosion time is given by the life time of the secondary plus the time for merging by GWR, the latter corresponding to the time scale for loss of angular momentum:

$$\frac{\dot{J}}{J} = -\frac{32}{5} \frac{G^3}{c^5} \frac{M_{WD1} M_{WD2} (M_{WD1} + M_{WD2})}{A_{ff}^4} \qquad (4)$$

(Landau & Lifshitz 1958) which, as we see, depends on the 8th power of $\alpha$.

A major uncertainty, most often overlooked, is that not every C+O WD plus C+O WD merging should give a SNe Ia. On the contrary, avoiding nondegenerate C ignition close to the surface of the more massive WD, either during the dynamical phase of merging or during the subsequent accretion from the massive disk, requires fine tuning (Mochkovitch & Livio 1989; Mochkovitch, Guerrero & Segretain, this volume).

We have not included other types of DD merging: O+Ne+Mg WD plus either O+Ne+Mg or C+O WD, C+O WD plus He WD, or He WD plus He WD. The cases involving O+Ne+Mg WDs would lead to accretion–induced collapse (AIC) rather than to SNe Ia (see Gutiérrez et al. 1996, and this volume). The He WD plus He WD case might produce detonation of the more massive WD, but that would only yield Fe–peak elements, in contrast with what is seen in SNe Ia spectra around maximum light. Merging of a C+O WD with a He WD might, in principle, produce a SN Ia either triggered by He detonation at the surface or by explosive C ignition at the center, but fine tuning in a still higher degree than for the double C+O WD case would be necessary to obtain a SN Ia through any of the two mechanisms.

It has been suggested that DD merging can also arise from systems which underwent only one CE episode (Webbink 1986). The first episode of mass transfer would take place by Roche lobe overflow. In this case the final separation of the systems would be wider and the SNe Ia rates from DD merging would change accordingly.

Hydrodynamic studies (Rasio & Livio 1996; Mochkovitch et al. 1996) should pronounce the final word on the plausibility of this scenario. An enrichment of C+O in the interstellar medium close to the SN, coming from the disrupted C+O WD, should be expected.

### 3.2. HE–STAR CATACLYSMICS (HECV)

In this scenario (Iben & Tutukov 1991, 1994), as in the preceding one, a first CE episode (when the primary is in the AGB phase) leaves a C+O WD plus a MS star in a close binary orbit. A second CE stage takes place while the secondary is in the red–giant (RG) phase. It produces a He star in a still closer orbit with the C+O WD. GWR later brings the system into contact and keeps it so. He is thus transferred steadily from the secondary to the primary and this can produce a SN Ia either by explosive ignition of the He layer accumulated on top of the C+O WD (sub–Ch SN Ia explosion) or by growth (as a consequence of steady burning of the accreted He into C and O) of the WD up to the Chandrasekhar mass, the issue depending on the mass–transfer rate $\dot{M}_{He}$.

The primordial primary mass range and the $q$ limits are similar to those for the DD scenario. The range of primordial separations $A_0$, however, is now bound from the conditions that the separation after the second CE stage has to be large enough to avoid immediate merging and small enough to allow GWR to bring the system into contact before completion of central He burning in the secondary: $A_{Roche} \leq A_0 \leq A_{He}$. Both conditions depend again on the value of the CE parameter $\alpha$.

For the rates found in this scenario, accumulation of He should lead to a *He detonation*, provided that the mass of the C+O WD is larger than $\simeq 0.6 - 0.7\ M_\odot$ (RBC95; Iben, this volume). As we will see in section 5 these systems are bad candidates to SNe Ia in ellipticals due to their short timescales to reach explosion (they could not survive up to the present time in systems in which star formation ceased several Gyrs ago).

### 3.3. CATACLYSMIC–LIKE SYSTEMS (CLS)

The CLS scenario in which the C+O WD accretes H by Roche lobe overflow from a MS or SG star has also been considered by Iben & Tutukov (1984, 1994). Van den Heuvel et al. (1992) have included those systems as a subclass of the so–called luminous supersoft X–ray sources. Rappaport, Di Stefano, & Smith (1994) have equally looked at them as possible progenitors of SNe Ia (see also Di Stefano et al., this volume). A CE episode produces a C+O WD plus a MS star in close orbit. Roche–lobe filling by the secondary leads to mass transfer on a thermal (Kelvin) time scale and to steady burning of the accreted H into He. This could produce a SN Ia either if He burns steadily into C+O, leading to a Ch explosion (Iben & Tutukov 1984; Rappaport, Di Stefano, & Smith 1994; CRB96), or if the He layer detonates and induces a second detonation in the (sub–Ch) C+O core (CRB96). It is expected to observe those systems associated with ionization nebulae (Di Stefano, Paerels & Rappaport 1995).

Primaries in the initial mass range $1.8 M_\odot \lesssim M_1 \lesssim 8 M_\odot$, plus secondaries with masses $M_2 \gtrsim 0.8 M_\odot$ would be the binary components at the origin. Roche–lobe filling takes place when the primary reaches the AGB stage. It reduces the separation between the two components of the system (once more depending on the value of the CE parameter $\alpha$), leaving a C+O WD as the residual of the primary.

Further conditions are required in order to ensure that mass transfer will take place unstably on a thermal time scale, that no second CE episode occurs, and that the mass–accretion rate is between the upper and the lower limit for stable thermonuclear burning of H into He. Roche–lobe filling by the secondary occurs due to its nuclear evolution.

The conditions for producing edge–lit detonations as SNe Ia are: (1) the WD mass must be $M_{WD} \gtrsim 0.7 M_\odot$, (2) a minimum mass $\Delta M_{He} \gtrsim 0.1 M_\odot$ must accumulate at the surface of the WD, and (3) the mass–accretion rate must be in the range $\dot{M}_{stable} \leq \dot{M} \leq \dot{M}_{upp}$, where $\dot{M}_{stable}$ is the lower limit for stable accumulation of He on top of the WD from accretion of H and $\dot{M}_{upp}$ is the upper limit for occurrence of He detonation, and is taken to be $\dot{M}_{upp} \simeq 5 \times 10^{-8} M_\odot \; yr^{-1}$ (Nomoto 1982). The two first conditions have also been adopted by Yungelson et al. (1995) for production of SNe Ia by WDs belonging to SS (see equally RBC95). For production of Ch explosions, conditions (1) and (2) change into: $M_{WD} + \Delta M = 1.4 M_\odot$. Condition (3) becomes: $\dot{M}_{upp} \leq \dot{M} \leq \dot{M}_{max}$, where $\dot{M}_{max}$ is the maximum accretion rate before the WD envelope expands to red–giant size.

An important uncertainty here is the value of $\dot{M}_{stable}$: should we take it just large enough to avoid nova–like outbursts, or to equally ensure the absence of large oscillations in the H burning rate and thus in the accreting WD radius (Iben 1982)? We will later see that adopting either of the two criteria significantly changes the predicted SNe Ia rates (see also CRB96). The value of $\dot{M}_{max}$, too, is uncertain. The earlier results of Nomoto, Nariai, & Sugimoto (1979) indicated that for accretion rates higher than $\sim 10^{-6} \; M_\odot \; yr^{-1}$ there was no static solution and the WD envelope expanded to RG size, which would form a CE. However, Hachisu et al. (1996) (see also Nomoto et al., this volume) now find a strong wind solution which would allow burning of H into He at a rate close to the former $\dot{M}_{max}$ even for much higher mass–transfer rates from the companion, the excess material being blown off by the wind. The time–dependence of the mass–accretion rate $\dot{M}$ will also be relevant to the rates of the SNe Ia arising from these systems (Di Stefano et al., this volume). A careful study of the surface burning process has been started by Hernanz et al. (this volume), and it is necessary for an accurate evaluation of the rates. In our simulations we find that, in the case of edge–lit detonations, low–mass WD explosions might be favored, since to avoid thermal pulses during surface H burning (which would prevent retention of the accreted material) in higher–mass WDs $\dot{M}$ should be too high to allow detonation.

### 3.4. CATACLYSMIC VARIABLES (CV)

CV are systems formed by a C+O WD plus a MS or SG star with a smaller mass. The WD results after formation of a common envelope during the AGB phase. The mass transfer occurs by Roche–lobe filling brought about and maintained by magnetic braking (Verbunt & Zwaan 1981) at a rate

given by:

$$\dot{M} = 1.9 \times 10^{-7} f^{-2} k^2 (M_{WD} + M_2) \frac{\mu^{5/3}}{7\mu - 4} \; M_\odot \; yr^{-1} \qquad (5)$$

where $\mu \equiv M_2/(M_{WD} + M_2)$, $f$ is deduced from the observed relationship between equatorial velocity and age in MS G stars, and $k$ from the relationship between angular momentum, mass, radius, and angular velocity appropriate for cool MS stars (Skumanich 1972; Smith 1979).

The CV which might produce SNe Ia would not be typical ones. The required mass–accretion rates are close to the upper limit of the range allowed for those systems. In some way those systems are an extension of the CLS considered above, but they differ from them in the value of q (here typical for CVs) and on the fact that mass transfer is due to the shrinkage of the orbit by magnetic braking rather than to nuclear evolution. The CV candidates to SNe Ia occupy a parameter space similar to that of the observed recurrent novae. These novae seem to be produced by a massive WD ($M_{WD} \geq 1.3 M_\odot$) and to have accretion rates $\dot{M} \geq 10^{-8}$ $M_\odot$ yr$^{-1}$. The recurrent episodes of nova explosion, however, do not seem to interfer with an overall gain in mass (MacDonald 1983). If accretion is sustained at rates $\dot{M} \lesssim \dot{M}_{upp}$ (see the previous subsection), a cold, degenerate He layer should form and when its mass becomes $M_{He} \gtrsim 0.1 M_\odot$ a He detonation might induce that of the C+O core. If the rates are $\dot{M} \gtrsim \dot{M}_{upp}$, the WD mass should continue to grow and a central C+O WD explosion would occur when the mass is close enough to the Chandrasekhar mass. Figure 3 and Figure 4 display the characteristics of those systems.

Therefore, a fraction of CV systems fulfill the conditions which could lead to SNe Ia. The question of the time dependence of the accretion rates is also relevant here. Setting the same conditions as in the CLS case, we have made a first estimate of the rates of SNe Ia in systems with different star formation histories and predicted SNe Ia statistics for them. As for HeCV, those systems could hardly produce explosions in galaxies where star formation stopped several Gyr ago.

3.5. SYMBIOTIC SYSTEMS (SS)

They were equally considered by Iben & Tutukov (1984), in the Ch explosion context. Recently, Munari & Renzini (1992), Kenyon et al. (1993), RBC95, and Yungelson et al. (1995) have looked at those systems as possible progenitors of sub–Ch SN Ia. In the simplest version of the SS scenario, the system remains always detached. Wind mass loss from the primary with ejection of a planetary nebula leads to the formation of a C+O WD. When the secondary, in turn, reaches the RG and AGB phases, a fraction of the wind that it emits can be accreted by the WD. H would be burned into He, and finally a He detonation could induce C detonation and produce a SNe Ia. Other evolutionary paths (which include either CE or semidetached phases) can also lead to the formation of a SS and, in a fraction of them, to a SNe Ia (Yungelson et al. 1995).

In a simple version of the SS scenario, the SNe Ia progenitors are systems with primaries in the initial mass range $0.8 M_\odot \lesssim M_1 \lesssim 5 M_\odot$ and orbital periods in the range $1\ yr \leq P \leq 10\ yr$. Within those ranges, the border between the systems evolving into SS and those evolving into cataclysmic variables (CV) is set by the mass ratio $q \simeq 0.26$ (Kenyon et al. 1993). The initial primary mass/WD mass relationship is taken from Weidemann (1986). The wind mass loss is treated by a Reimers–type parameterization (Reimers 1975):

$$\dot{M}_1 = -4 \times 10^{-13} \eta \frac{R_1 L_1}{M_1}\ M_\odot\ yr^{-1} \qquad (6)$$

(see RBC95). Further refinements may include separation increase due to mass loss from the system by the winds, the possibility of either a CE or a semidetached stage, the efficiency of mass retention, and the requirement of stable burning regime of the accreted H (Yungelson et al. 1995). We will see, however, that their inclusion is mostly equivalent to the introduction of an efficiency factor $\eta_{SS}$.

The wind accretion can be considered as a case of the Bondi–Hoyle accretion process (Bondi & Hoyle 1944):

$$\dot{M}_2 = -\left[\frac{GM_2}{V_{wind}^2}\right]^2 \left[\frac{1}{1 + (V_{orb}/V_{wind})^2}\right]^{3/2} \frac{\alpha_{accr}^2}{2 A^2} \dot{M}_1 \qquad (7)$$

where $\alpha_{accr}$ is a numerical constant between 1 and 2 which comes from the approximations made when deriving the amount of mass accreted during the interval $\Delta t$ from the relative velocity of the accreting component with respect to the matter flowing out (Boyle 1984; Boffin & Jorissen 1988). A is the orbital separation, $V_{wind}$ is the wind velocity, and $V_{orb}$ is the orbital velocity of the accreting WD. The accretion process is dominated by the collisions between the streams of gas flowing in opposite directions behind (with respect to the mass–losing star) the accreting WD.

## 4. The Basic Input

### 4.1. THE INITIAL MASS FUNCTION

Since there are no clear indications that the stellar initial mass function (IMF) significantly depends on the environment, we adopt an IMF in-dependent of time and of galaxy type. For the IMF of the primaries in the binary systems involved in the different scenarios, several expressions have been adopted. Iben & Tutukov (1984) used the Salpeter (1955) form $dN \propto M^{-2.5} dM$. In our models we have adopted the IMF from Scalo (1986):

$$dN \propto exp[-1.09(log\ M + 1.02)^2]d(log\ M) \qquad (8)$$

where $M$ is the primary mass. We have afterwards compared the results obtained by using (8) with those corresponding to the IMF of Salpeter (1955) (see below, and RBC95).

### 4.2. THE STAR FORMATION RATES

The star formation rates (SFR) required to compute SNe Ia production in each scenario can be taken from detailed models for the evolution of galaxies. SFR for the main galaxy types have been derived by Sandage (1986). The rates obtained from chemodynamical simulations (Hensler & Burkert 1990) seem to confirm their general features.

Sandage's (1986) SFRs corresponding to the different galaxy types, from E0 through Sd, were devised to reproduce the variation of stellar populations with morphological type deduced by Gallagher et al. (1984). In the case of elliptical galaxies, for instance, a burst of star formation happens at the beginning and lasts for only a couple of Gyr. SFR as high as $\sim 5 \times 10^{-9} M_\odot\ yr^{-1}$ are reached during $\sim 2 \times 10^8 - 1 \times 10^9\ yr$, which is $\sim 50$ times the current SFR in the solar neighbourhood. The different time evolutions of those SFR allow to explain the bulge/disk ratio as a function of Hubble type, the variation in disk surface brightness along the Hubble sequence, the integrated colors (which change monotonically as a function of galaxy type), the mean ages of the disks along the sequence, and that, at present, the SFR per unit mass in Sc galaxies is much higher than in S0 and Sa galaxies. We have adopted for our estimates the SFR of Sandage (1986) for the different Hubble types.

### 4.3. THE DISTRIBUTION OF MASS RATIOS AND PRIMORDIAL SEPARATIONS

It is not clear that a general distribution function $F(q, A, M_1)$ independent from the environment exists, nor can it be taken for granted that $f(q, P)$ ($P$ being the orbital period) can be decomposed into $f(q, P) = f_1(q)f_2(P)$ as we will assume here. The most complete observational study of those distributions has been made by Duquennoy & Major (1991), for solar–type binaries. We have assumed that half of the stars born in the galaxies are in binaries and explored the effects of adopting different distributions for

$q$ and $P$. We have first adopted the distribution of Duquennoy & Mayor (1991):

$$f(q) \propto exp\left[-\frac{(q-0.23)^2}{0.35}\right] dq \qquad (9)$$

assuming that it can be extrapolated up to masses $M \lesssim 8M_\odot$. We have also tried, as in Iben & Tutukov (1984), the distribution of mass ratios proposed by Popova et al. (1982): $f(q) = Cq^\alpha$ (where $C$ satisfies the normalization condition that $\int_0^1 f(q)dq = 1$), with $\alpha = 1$, and the one suggested by Mazeh et al. (1992).

Different choices are also possible for the distribution of primordial separations. We use again the distribution given by Duquennoy & Mayor (1991):

$$dN \propto exp\left[-\frac{(log\ P - 4.8)^2}{10.58}\right] d(log\ P) \qquad (10)$$

with $P$ in days. From this we immediately derive the distribution of primordial separations $A_0$ using Kepler's law. Here we have also tried the "flat" distribution of Popova et al. (1982): $dN \propto d\ log\ A_0$.

We generate an ensemble of systems by Monte Carlo trials, with probabilities given by the IMF of the primaries and the $q$ and $A_0$ distributions, and use a scenario code to decide the fate of each system (RBC95, CRB96).

### 4.4. STELLAR LIFETIMES

The lifetimes of the secondaries in the different nuclear burning stages enter into the determination of the epoch of SNe Ia explosion. We adopt the results from Schaller et al. (1992), which include the effects of mass loss via stellar wind. We track the internal structures of both the donors and the WD progenitors, at any given time, by means of the diagrams of Politano (1988) and de Kool (1992).

### 4.5. THE ROLE OF METALLICITY

The question is often raised of whether changes in SNe Ia characteristics might mainly reflect differences in the metallicities of the stellar populations involved. We must note here that although early–type galaxies are, in general, more metal–rich than late–type galaxies, the trend is rather modest: the whole range spans a factor $\lesssim 3-5$ (Roberts & Haynes 1994). The global metallicity of a galaxy depends more on its mass than on its Hubble type. In this way, since early–type galaxies are generally more massive than late–type galaxies, the trend is reinforced. Differences in metallicity, however, can only act through the factors which determine SNe Ia rates and characteristics.

Differences within the above range are not likely to change the IMF of the primordial primaries in the progenitor systems of the SNe Ia. It is even less likely that they would affect the primordial $q$ and $A_0$ distributions. Somewhat more significant are differences in stellar lifetimes which result from metallicity differences. They are taken into account in our calculations according to the results of Schaller et al. (1992).

A possible influence of metallicity on SNe Ia might be through the delay of explosive C ignition by the convective Urca process. Higher metallicity might mean a more effective cooling and thus a higher ignition density. That might, for instance, favor the delayed pulsating detonations with small $^{56}$Ni production. In this way, the average higher metallicity of early type galaxies could be connected with dimmer explosions. However, this effect has not been properly evaluated and it might be minor.

### 4.6. RATES AT HIGH REDSHIFTS

The fact that the dominant stellar populations within each morphological type do change with redshift due to cosmic evolution will also affect the rates and characteristics of SNe Ia explosions if, as suggested by the observations, they basically depend on the age of the populations involved.

Reaching up to redshifts $z > 0.3$ in the statistics of SNe Ia corresponds to have a lookback time of a few Gyr into the endpoint of binary evolution leading to SNe Ia, and it is interesting to see how the frequency and the average properties of the explosions might vary, since different scenarios predict different dependences of the rates and other characteristics on redshift (RBC95; CRB96). In order to predict rates as a function of redshift, we must first adopt some model of galaxy formation. In Table 1 we show the typical redshifts at which galaxies with $M \sim 10^{11}\ M_\odot$ should likely form, depending on the cosmological model. We give the values corresponding to a cosmological $1\sigma$ density fluctuation for protogalaxies to form. We can assume that star formation begins as soon as the protogalaxy collapses. There is not so much known about the delay between the start of the collapse of the protogalaxy and the onset of star formation, and several episodes of star formation might occur along the collapse process of the protogalaxy. However, this time interval (given by the characteristic free–fall time) should be much shorter than the life times of the SNe Ia progenitors.

Morphological segregation tends to place a larger number of ellipticals in the central regions of the galaxy clusters. The evolution of the SNe Ia rates in ellipticals can be followed very well in searches up to redshifts $z \sim 0.5$. Since this corresponds to the a lookback time of $\sim 6$ Gyr, the SNIa rates might be significantly enhanced as compared with the present epoch for some scenarios. The systems undergoing explosion in those more distant galaxies being younger than in nearby galaxies, there is a way to test the different SNe Ia scenarios. Ellipticals present the cleaner look at the variation of the properties of SNe Ia with age of progenitor, since the population should be less mixed than in spirals: star formation seems to have stopped several Gyrs ago. For sub–Ch explosions the average mass of the WD should be larger in far–away ellipticals galaxies and lead to more luminous SNe Ia as compared to SNeIa in ellipticals at redshift $z = 0$ (RBC95). This provides an excellent opportunity to determine whether sub–Ch or Ch explosions are dominant. In the case of Ch explosions, no increase in luminosity is predicted in ellipticals at $z = 0.5$ as compared with present time ones. Also, the $\Delta m_{15}$ parameter should decrease at larger redshifts if a sub–Ch scenario is at work. In general, if the sub–Ch scenarios were correct, a clear tendency to brighter SNe Ia when going towards distant galaxies should be seen. The masses of the exploding WDs should moderately grow from $z = 0$ to $z = 0.5$, and in sub–Ch type of explosions a small increase in mass of the WD gives a substantial increase in the mass of $^{56}$Ni synthesized in the explosion (Ruiz–Lapuente et al. 1993; Woosley & Weaver 1994; Livne & Arnett 1995). The average exploding WD masses at those (already surveyed) redshifts would be around 0.9 $M_\odot$ and the explosions could produce as much as 0.5 $M_\odot$ of $^{56}$Ni. These SNe Ia would look "normal". Galaxies at redshifts twice the presently surveyed would host very bright SNe Ia as a result of the explosion of WDs of $\simeq 1.0$–$1.2$ $M_\odot$. Are those massive sub–Ch WDs expected in galaxies where star formation is still going on? The average mass of the progenitor systems is larger in those galaxies as compared with present–time ellipticals and in the latest types such massive sub–Ch WDs should still be exploding now. In the case of Ch explosions, in contrast, no increase in luminosity, along the Hubble sequence would be expected, unless the metallicity effect mentioned in the previous section were at work.

TABLE 1. Redshifts at which galaxies of a given mass form, for different cosmological scenarios and parameters

|  | CDM ($h=0.5$, $\Omega_\Lambda=0$, $\Omega_M=1$) | CDM ($h=0.5$, $\Omega_\Lambda=0.6$, $\Omega_M=0.4$) | MDM ($h=0.5$, $\Omega_\Lambda=0$, $\Omega_M=1$) | CDM ($h=0.75$, $\Omega_\Lambda=0$, $\Omega_M=1$) | CDM ($h=0.7$, $\Omega_\Lambda=0.6$, $\Omega_M=0.4$) |
|---|---|---|---|---|---|
| $M \sim 10^{11} M_\odot$ ($1\sigma$) | 2.9 | 2.6 | 0.7 | 3.8 | 1.7 |

## 5. SNe Ia: statistics and time evolution

Table 2 summarizes the results obtained for different scenarios and galaxy types. The rates at t = 10 Gyr are given in the third, fourth, and fifth columns of the Table, for E0, Sb, and Sc galaxies, respectively, and for the four different types of systems: CLS, SS, HeCV, and DD. For SS, both the rates from the model of RBC95 ($\eta_{SS} = 1$) and from Yungelson et al. (1995) are shown. In late–type galaxies, the rates from any scenario (with one exception) are higher (by factors of 2–5) than in early–type galaxies. That is close to the trend found by Della Valle & Livio (1994).

We expect at the present time, in late–type galaxies, the SNe Ia rates from SS and from He cataclysmics to be larger and those from DD merging to be similar to the rates from sub–Ch CLS (for the most restrictive $\dot{M}_{stable}$, and if we assumed a 100% efficiency in SNe Ia production from DD merging), while CLS would give the dominant contribution to Ch explosions. For the less restrictive bound on $\dot{M}_{stable}$, only the maximal rates from SS would be larger than those from CLS (both sub–Ch and Ch). Note that, in this case, the rates from sub–Ch SNe Ia in CLS would be constant when going from early–type galaxies to late–type galaxies. Dominance of sub–Ch explosions, as in early–type galaxies, appears likely in late–type galaxies, but a mixture with Ch explosions would be possible even for $\alpha < 1$ (the latter mainly coming from CLS). Also, the average mass of the exploding WDs (and thus the luminosity of the SNe Ia and the velocities of the ejecta) should be higher, for sub–Ch explosions, than in early–type galaxies.

Concerning the SS scenario, note that the result from the detailed modeling by Yungelson et al. (1995) of the galactic population of SS with WD accretors is basically equivalent, for the SNe Ia rates, to assign a value $\eta_{SS} \simeq 0.05$ to the efficiency factor for SNe Ia production in the version of the same scenario adopted in RBC95. It is also seen in Table 2 that SNe Ia from CVs would not be found in E galaxies. The timescale to explosion, which adds that for evolution of the primary to the typical timescale for Roche–lobe filling of the secondary and transfer of mass by magnetic braking, is too short to allow explosions of WDs in an old population.

In Figures 1 and 2 we show the results for an instantaneous burst of star formation, and for the different scenarios described in Section 3. The results for SS, HeCV, and DD merging are displayed in Figure 1, and those for CLS (sub–Ch explosions) are given in Figure 2. These results can also be taken as a first approximation to the evolution of the SNe Ia rates in an elliptical galaxy. In the second and third panels of Figure 1, we show the time evolutions of the average mass of the exploding WD and of the average velocity of the ejecta, respectively, for the SS scenario. As discussed in RBC95, the decrease in the average WD mass would correspond to also decreasing $^{56}$Ni masses and thus to dimmer SNe Ia as the time since the initial burst increases. The same trend is found for the HeCV scenario, but as it can be seen in the third panel of Figure 1, no SNe Ia should be expected from this scenario in early–tipe galaxies (the different curves, in this panel and also in the next one, correspond to different values of the CE parameter, $\alpha$). The fourth panel shows the DD case: C+O plus C+O WD mergings can still be happening at present, even in early–type galaxies, if $\alpha$ is not significantly smaller than 1, but for this scenario (Ch explosions) no time evolution of the explosion characteristics (brightness, expansion velocities) is predicted (see, however, Canal, this volume, on the possible effects of cooling).

TABLE 2. SNe Ia Rates at t = 10 Gyr[1]

| Scenario | $\alpha$ | Galaxy Type | | |
|---|---|---|---|---|
| | | E0 | Sb | Sc |
| CLS[a] | 1 | **$2.0\times10^{-5}$** | **$1.1\times10^{-4}$** | **$1.1\times10^{-4}$** |
| | | 0.02 | 0.03 | 0.02 |
| CLS[b] | 1 | **$4.6\times10^{-4}$** | **$4.6\times10^{-4}$** | **$4.6\times10^{-4}$** |
| | | 0.37 | 0.11 | 0.07 |
| CV[a] | 1 | **0** | **$3.9\times10^{-5}$** | **$6.0\times10^{-5}$** |
| | | 0 | 0.01 | 0.01 |
| Symbiotics[c] | 1 | **$1.4\times10^{-3}$** | **$4.8\times10^{-3}$** | **$4.8\times10^{-3}$** |
| | | 1.12 | 1.20 | 0.72 |
| Symbiotics[d] | 1 | **$7.0\times10^{-5}$** | **$2.7\times10^{-4}$** | **$3.2\times10^{-4}$** |
| | | 0.06 | 0.07 | 0.05 |
| He cataclys. | 1 | **0** | **$2.3\times10^{-4}$** | **$4.6\times10^{-4}$** |
| | | 0 | 0.06 | 0.07 |
| He cataclys. | 0.3 | **0** | **$4.6\times10^{-4}$** | **$9.1\times10^{-4}$** |
| | | 0 | 0.11 | 0.14 |
| DD[e] | 1 | **$7.0\times10^{-5}$** | **$1.6\times10^{-4}$** | **$2.0\times10^{-4}$** |
| | | 0.06 | 0.04 | 0.03 |
| DD[e] | 0.3 | **0** | **$1.4\times10^{-4}$** | **$1.8\times10^{-4}$** |
| | | 0 | 0.03 | 0.03 |
| CLS[e] | 1 | **0** | **$4.6\times10^{-4}$** | **$4.6\times10^{-4}$** |
| | | 0 | 0.11 | 0.07 |
| CV[e] | 1 | **0** | **$1.6\times10^{-4}$** | **$2.5\times10^{-4}$** |
| | | 0 | 0.04 | 0.04 |

[1] Boldface figures are rates per galaxy of $10^{10}$ $M_\odot$. Roman figures are rates in SNU
[a] More restrictive bound on $\dot{M}_{stable}$
[b] Less restrictive bound on $\dot{M}_{stable}$
[c] From RBC95
[d] From Yungelson et al. (1995)
[e] Chandrasekhar–mass explosions

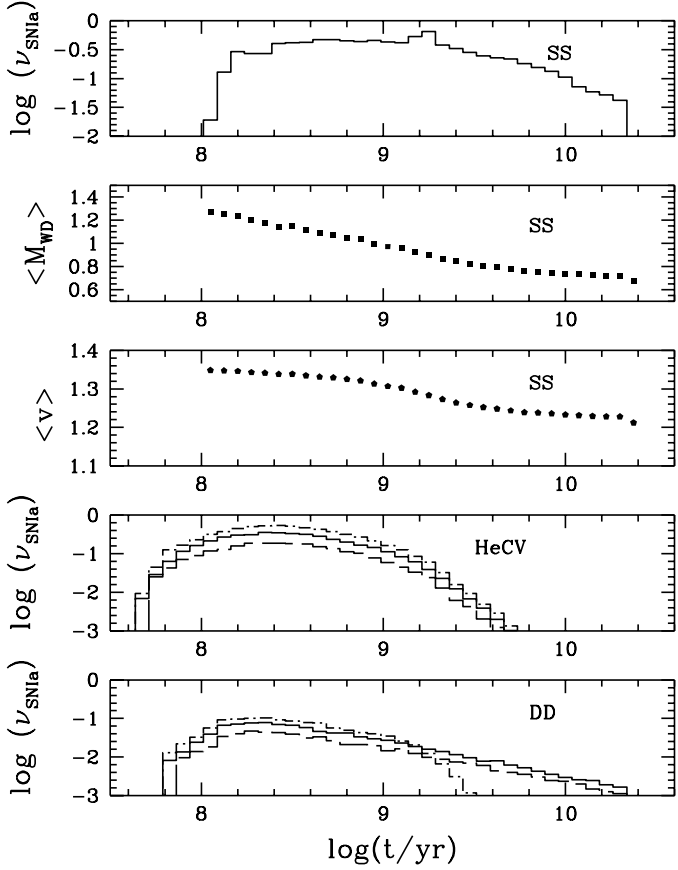

Figure 1. Time evolution of the SNe Ia rate and characteristics after an instaneous burst of star formation, for three different scenarios: SS, HeCV, and DD merging. Shown in the second and third panels are the time evolutions of the average mass of the explosing WD and of the average explosion energy, respectively, for the SS scenario. In the fourth and fifth panels, the continuous, dashed, and dot–dashed histograms respectively correspond to the values 1, 0.5, and 2 of the CE parameter $\alpha$ (see text)

In the first panel of Figure 2 we see, for CLS (sub–Ch explosions), the dependence of the result on the values adopted for $\dot{M}_{stable}$ (H burning) and for the upper limit for He detonation to occur ($\dot{M}_{upp}$): the dot–dashed histogram corresponds to the $\dot{M}_{stable}$ which is just high enough to avoid

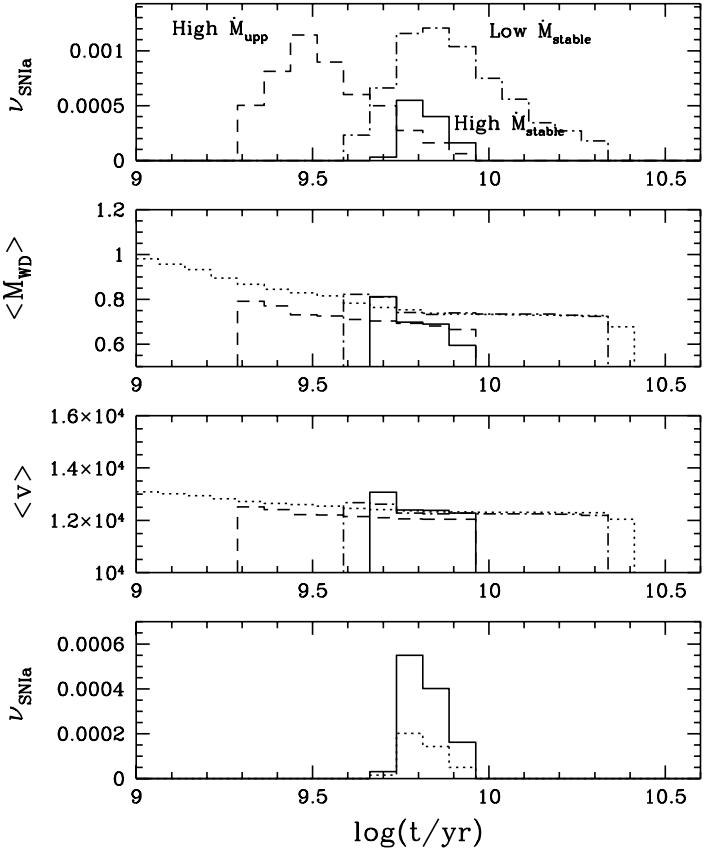

*Figure 2.* Time evolution of the SNe Ia rate and characteristics after an instaneous burst of star formation, for the CLS scenario (sub–Ch explosions). In the first panel, the three histograms corresponds to two different choices for $\dot{M}_{stable}$ and $\dot{M}_{upp}$ (see text). Shown in the second and third panels are the time evolutions of the average mass of the exploding WD and of the average explosion energy, respectively, also for the three cases (dotted line corresponds to the SS scenario). In the fourth panel, our most "standard" result is compared with that obtained by adopting the same IMF and $q$ and $A_0$ distributions as Iben & Tutukov (1984) and Tutukov & Yungelson (1994)

explosive burning of the accreted H, whilst the solid histogram (continuous line) is the result for a $\dot{M}_{stable}$ high enough to ensure steady H burning (Iben 1982). We see that, in the latter case, no SNe Ia would be produced at present, for this scenario, in early–type galaxies. The dashed histogram corresponds to the combination of the "standard" (high) value for $\dot{M}_{stable}$ with $\dot{M}_{upp} = 10^{-7}\ M_\odot\ yr^{-1}$, which is twice the value given by Nomoto (1982). Shown in the second and third panels of Figure 2 are the respective time evolutions of the average mass of the exploding WD and of the average velocities of the ejecta (for the higher $\dot{M}_{stable}$ and the "standard" $\dot{M}_{upp}$). We see that the range of variation of SN characteristics would be very narrow and that the average mass of the exploding WD, in elliptical galaxies, should be $M_{WD} \sim 0.7 - 0.8 M_\odot$. In the bottom panel of Figure 2 we see the effect of adopting the IMF and the $q$ and $A_0$ distributions used by Iben & Tutukov (1984) and Tutukov & Yungelson (1994).

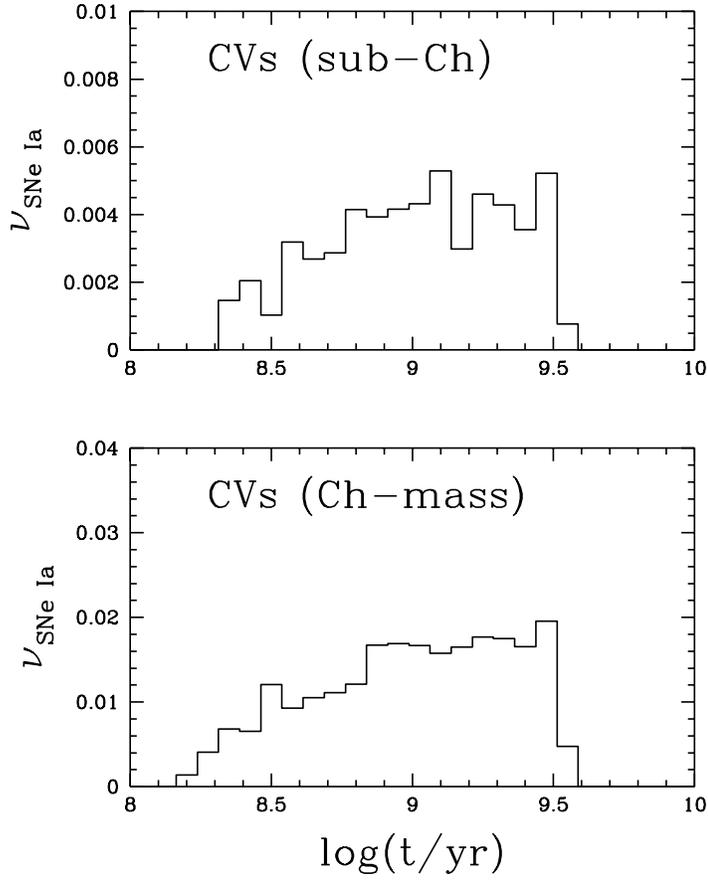

Figure 3. Time evolution of the rates after an instantaneous burst of star formation, for the CV scenario. Both sub–Chandrasekhar and Chandrasekhar explosions are considered. According to this scenario SNe Ia would not start to explode before $10^{8.5}$ yr and would stop around $10^{9.5}$ yr after the birth of the progenitor systems.

Figure 3 shows the time evolution of the SNe Ia rates from CV systems. The distributions of masses of the accreting WDs at the start of the ac-cretion process and that of the orbital periods of the systems are shown in Figure 4.

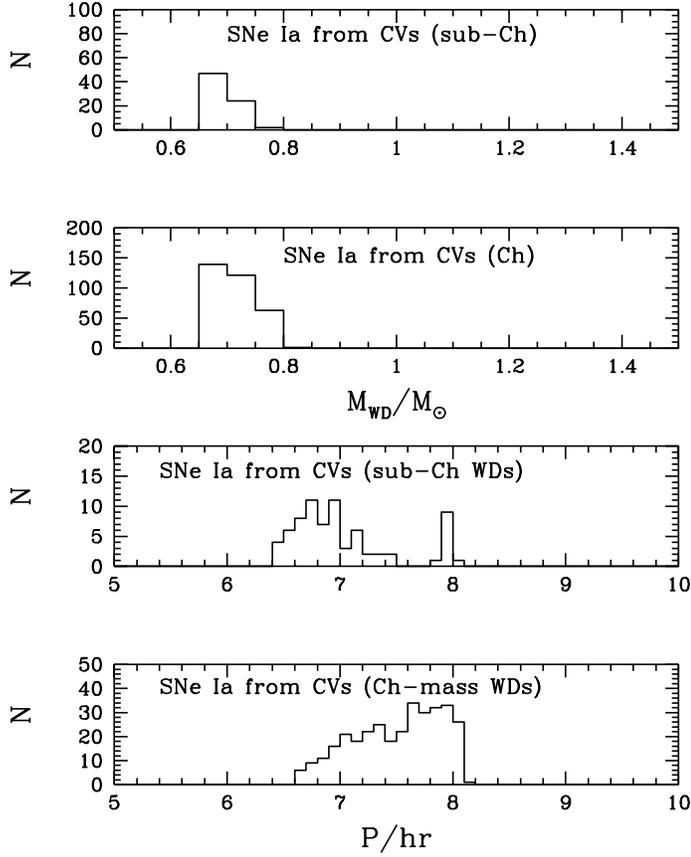

*Figure 4.* Distribution of SNe Ia progenitors from CV systems in orbital period of the system and in mass of the exploding WD at the start of mass accretion, both for Chandrasekhar and sub–Chandrasekhar explosions.

The variation of the SNe Ia rates as a function of redshift, assuming that an instantaneous burst of star formation took place at $z \sim 5$, are shown in the first panel of Figure 5 for the DD and HeCV scenarios, and in the second panel for the SS scenario. The effects of adopting different IMF and $q$ and $A_0$ distributions is also shown there. In the last panel we compare the three scenarios for $0 \leq z \leq 1$ (for the SS scenario, we show both the evolution after an instantaneous burst and for the SFR history corresponding to an E0 galaxy).

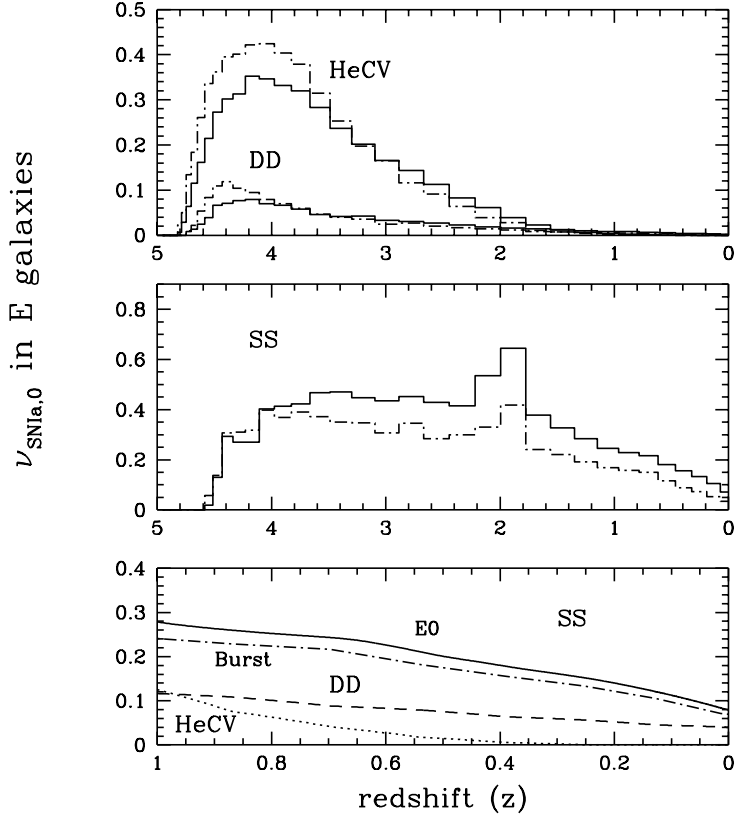

*Figure 5.* The SNe Ia rates from an instantaneous burst of star formation, for the same three scenarios of Figure 1, in redshift space. In the first and second panels, two different choices for the IMF and the $q$ and $A_0$ distributions are compared (see text). In the third panel the results for the three scenarios are compared for $z \leq 1$. Dashed line corresponds to the DD scenario, dotted line to the HeCV scenario, and dot–dashed line to the SS scenario. The case of an E0 galaxy (SS) is shown by the continuous line. To allow comparison, the absolute scale for the DD and HeCV scenarios has been shifted upwards

As it has been addressed above, there is a double dependence in the evolution of SNe Ia rates and characteristics as a function of redshift: both on the scenario and on the Hubble type of the parent galaxy. That opens up the possibility of discriminating among scenarios on basis to SNe Ia statistics at high enough redshifts. If we consider, for instance, elliptical galaxies only, we have that the SNe Ia rates would rise faster for the SS scenario than for DD merging. For any scenario, the rates should increase

## CANDIDATES TO SNe Ia

| | | | |
|---|---|---|---|
| **DD** | • Chandrasekhar C+O | | hardly in E galaxies if $\alpha < 1$ |
| **HeCV** | • Edge–lit detonation (sub–Ch) | | not in E galaxies |
| **CLS** | • Chandrasekhar C+O | $\Longrightarrow$ | not in E galaxies (high $\dot{M}$ requires massive systems) |
| | • Edge–lit detonation (sub–Ch) | $\Longrightarrow$ | in both S and E low–mass sub–Ch |
| **CV** | • Chandrasekhar C+O | $\Longrightarrow$ | in S but not in E galaxies |
| | • Edge–lit detonation (sub–Ch) | $\Longrightarrow$ | only low–mass sub–Ch |
| **SS** | • Edge–lit detonation (sub–Ch) | | wide range of sub–Ch |

*Figure 6.* Summary on the possible candidates to SNe Ia as concluded from our results

less steeply with $z$ in spirals than in ellipticals. This is due to the fact that in late–type galaxies the explosions are due, at any time, to a mixture of systems which spans a broad interval of ages: the smaller numbers of old systems as the evolution proceeds is partially compensated by the contribution from younger systems.

If we include the possibility of sub–Ch WD explosions, then the time dependence of the average mass of the exploding WD produces a evolution in time of the average $^{56}$Ni mass and explosion energy. In this way, fainter SNe Ia, with lower expansion velocities and faster light curve de-clines (lower WD masses) would dominate in early–type galaxies, where star formation basically stopped several Gyr ago, whilst the more massive WDs associated with younger systems (brighter explosions, higher expansion velocities, slower light curve declines) would only explode in late–type galaxies, where star formation is still going on. In case of a mixture of Ch and sub–Ch explosions, the former would be more common in late–type than in early–type galaxies and this would reinforce the trend to brighter, more energetic, and more slowly declining average SNe Ia along the Hubble sequence from E0 to Sd.

To summarize the previous results (see also Figure 6):

a) *HeCV* cannot be the progenitors of SNe Ia in elliptical galaxies.

b) *DD mergings* (C+O WD plus C+O WD) might contribute to the SNe Ia rate in early–type galaxies, provided that in CE episodes the efficiency with which orbital energy is used to eject the envelope is not too low (CE $\alpha \simeq 1$). For this scenario (Ch explosions always) the origin of any possible correlation of SNe Ia characteristics with Hubble type of the parent galaxy remains to be investigated (the metallicity effect previously mentioned still lacks any quantitative evaluation, as well as the possible effects of cooling). The coalescence of WDs should be further studied by hydrodynamic simulations, in order to ascertain its outcome.

c) *SS* would explain the observed correlations: brighter SNe Ia and higher expansion velocities in galaxies of later types, together with the peak brightness/rate of decline relationship, by the explosion of WDs with different masses (sub–Ch explosions). After the end of star formation in early–type galaxies, the average mass of the exploding WD would monotonously decrease, being $\sim 0.8 M_\odot$ at present in ellipticals, with narrow dispersion. In late–type galaxies, the continuous process of star formation would produce a spread in the exploding WD masses, the average mass being larger, at present, than in early–type galaxies. From this scenario we would expect a whole range of sub–Chandrasekhar explosions to be seen. Detail calculations of wind accretion are needed. The accreting systems prior to explosion might show up in the X–ray domain.

d) *CLS* (sub–Ch explosions) would also give a narrow range of explosion characteristics in elliptical galaxies, but the average mass of the exploding WD would always be $M_{WD} \sim 0.7 - 0.8 M_\odot$. On the other hand, there should not be any Ch explosions coming from those systems in early–type galaxies at present. A dominance of Ch explosions in Sb–Sc is expected. Thus, if CLS were giving the dominant sub–Ch contribution (together with Ch explosions), the distribution of SNe Ia characteristics (peak luminosities, expansion velocities) should appear close to bimodal rather than spanning a continuous range. These systems would show up as luminous supersoft X–ray sources: those leading to sub–Ch explosions should appear as less luminous than the Ch ones and with a softer spectrum.

d) *CV*. Concerning SNe Ia arising from CV, the allowed region of masses of the components, separations, and accretion rates would place them in the same region of the parameter space as recurrent novae. Such systems should also be observable as X-ray sources (another clase of supersoft sources). CV–SNe Ia are not expected to be found in ellipticals, since they have an evolutionary time scale too short.

Detection of H in the early or late–time spectra would clearly favor single–degenerate progenitor systems with either MS, RG, or AGB nondegenerate components.

Are the ages of the SNe Ia progenitors of the order of $5 \times 10^8$ yr, $10^9$ yr, $5 \times 10^9$ yr, or $10^{10}$ yr? Are the systems exploding in elliptical and in spiral galaxies different? It seems clear that setting a narrow band for the ages of SNe Ia progenitors, should give us a clearer indication on the nature of the corresponding systems.

The authors acknowledge financial support from the DGICYT of Spain. P.R.L would like to thank M. de Kool and the members of the binaries group of the Max–Planck Institut für Astrophysik for useful conversations.